\documentclass[aps,pre,twocolumn,showpacs,superscriptaddress]{revtex4-1}
\usepackage{graphicx}  % needed for figures
\usepackage{dcolumn}   % needed for some tables
\usepackage{bm}        % for math
\usepackage{amssymb}   % for math
\usepackage[final]{epsfig}
\usepackage{amssymb}
\usepackage{latexsym}
\usepackage{wrapfig}
\usepackage{amsmath}
\usepackage{booktabs}
\usepackage{threeparttable}
\usepackage{xcolor}
\usepackage{amsthm}
\usepackage{lipsum}
\usepackage{hyperref}
\usepackage{enumitem}
\newcommand{\bfa}{{\bf a}}

\newcommand{\bfe}{{\bf e}}

\newcommand{\bfh}{{\bf h}}

\newcommand{\bfn}{{\bf n}}
\newcommand{\bfo}{{\bf o}}
\newcommand{\bfp}{{\bf p}}
\newcommand{\bfq}{{\bf q}}
\newcommand{\bfr}{{\bf r}}

\newcommand{\bft}{{\bf t}}

\newcommand{\bfx}{{\bf x}}
\newcommand{\bfy}{{\bf y}}

\newcommand{\bfI}{{\bf I}}

\newcommand{\bfQ}{{\bf Q}}
\newcommand{\bfR}{{\bf R}}

\newcommand{\bfU}{{\bf U}}

\newcommand{\beq}{\begin{equation}}
\newcommand{\eeq}{\end{equation}}
\newcommand{\beqs}{\begin{eqnarray}}
\newcommand{\eeqs}{\end{eqnarray}}

\newcommand{\calD}{{\cal D}}

\newcommand{\calU}{{\cal U}}

\begin{document}

% The following information is for internal review, please remove them for submission
%\widetext
%\leftline{Version xx as of \today}
%\leftline{Primary authors: Joe E. Physics}
%\leftline{To be submitted to (PRL, PRD-RC, PRD, PLB; choose one.)}
%\leftline{Comment to {\tt d0-run2eb-nnn@fnal.gov} by xxx, yyy}
%\centerline{\em D\O\ INTERNAL DOCUMENT -- NOT FOR PUBLIC DISTRIBUTION}

% the following line is for submission, including submission to the arXiv!!
%\hspace{5.2in} \mbox{Fermilab-Pub-04/xxx-E}

\title{Evolving, complex topography from combining centres of Gaussian curvature}
%\input author_list.tex       % D0 authors (remove the first 3 lines
                             % of this file prior to submission, they
                             % contain a time stamp for the authorlist)
                             % (includes institutions and visitors)
\author{Fan Feng}
\affiliation{Cavendish Laboratory, University of Cambridge, Cambridge CB3 0HE, UK}
\author{John S. Biggins}
\affiliation{Department of Engineering, University of Cambridge, Cambridge CB2 1PZ, UK}
\author{Mark Warner}
\email{mw141@cam.ac.uk}
\affiliation{Cavendish Laboratory, University of Cambridge, Cambridge CB3 0HE, UK}

\date{\today}

\begin{abstract}
Liquid crystal elastomers and glasses can have significant shape change determined by their director patterns.
Cones deformed from circular director patterns have non-trivial Gaussian curvature localised at tips, curved interfaces, and intersections of interfaces.
We employ a generalised metric compatibility condition to
characterize two families of interfaces between circular director patterns -- hyperbolic and elliptical interfaces, and find that the deformed interfaces are geometrically compatible. We focus on hyperbolic  interfaces to design complex topographies and non-isometric origami, including n-fold intersections, symmetric and irregular tilings. The large design space of three-fold and four-fold tiling is utilized to quantitatively inverse design an array of pixels to display target images. Taken together, our findings provide comprehensive design principles for the design of actuators, displays, and soft robotics in liquid crystal elastomers and glasses.
\end{abstract}

%\pacs{}
\maketitle

%\section{\label{sec:level1}First-level heading}
% sections are not used for PRL papers
\section{Introduction}
Liquid crystal elastomers (LCEs) and glasses are solids that can change length by between 10--400\%  along their ordering direction, their director $\bfn$. These reversible length changes can be driven by heat, light or solvent. Even more remarkably, their director fields and hence their mechanical response can be programmed spatially to give a non-uniformity that changes the Gaussian curvature on actuation since the metric changes in a non-trivial manner. Impeding such curvature changes leads to stretch away from the new relaxed state and thus to strong mechanical response, in contrast to that encountered in bend and other such isometries. For example,  flat sheets programmed with arrays of +1 nematic defects actuate into arrays of cones, and, experimentally have been shown to lift thousands of times their own mass through strokes that are hundreds of times their own thickness \cite{ware2015voxelated,white2015programmable}.This geometry-driven paradigm has been termed ``metric mechanics" (see the review \cite{warner2019topographic}) and is an underlying motivation of this study of complex, evolving topographies.

\begin{figure}[!th]
	\centering
	\includegraphics[width=\columnwidth]{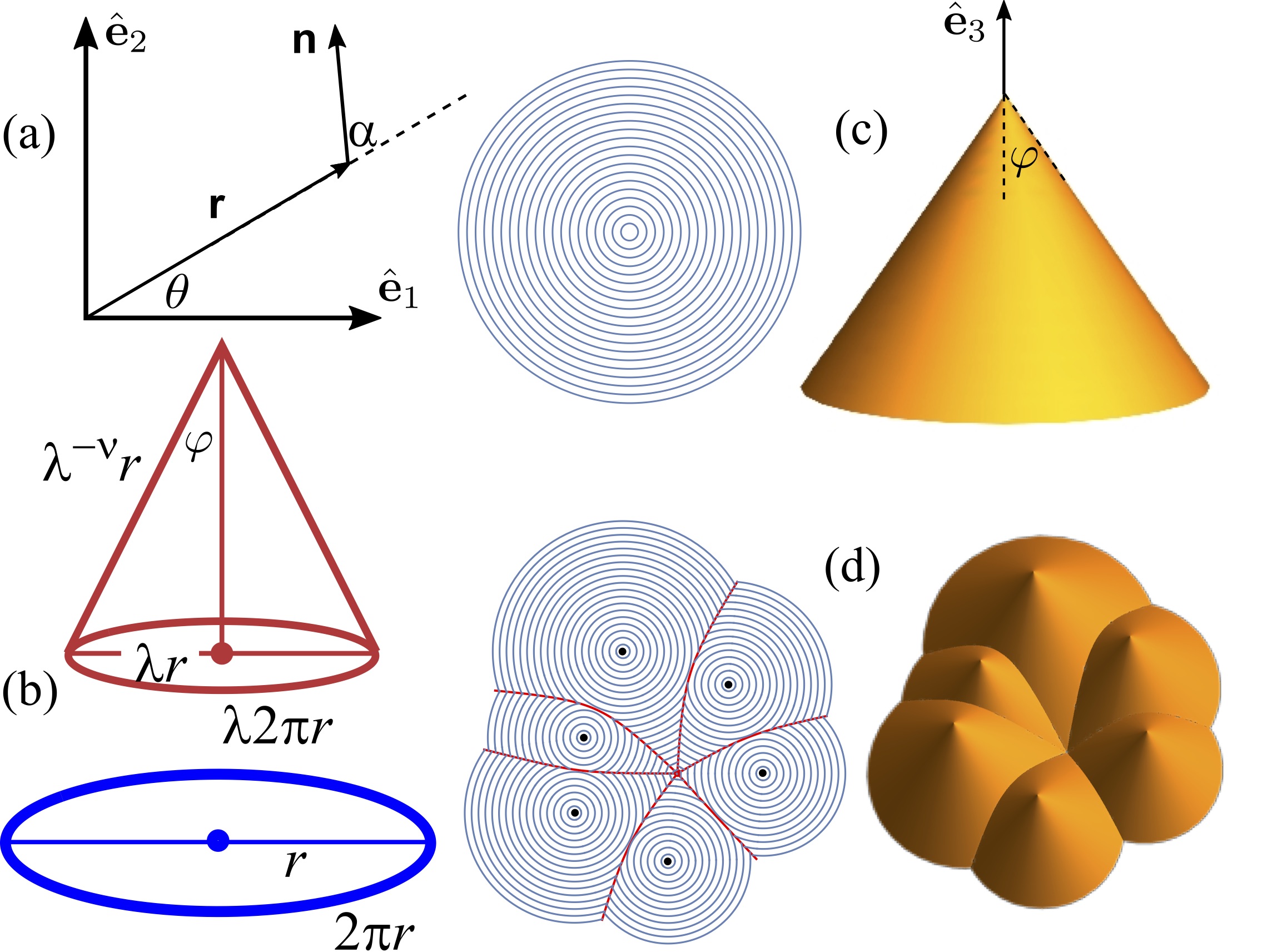}
	\caption{(a) A director field $\bfn(\bfr)$ making an angle $\alpha$ to the radial vector $\bfr$ itself at angle $\theta$. (b) A circumference $2\pi r$ contacts by a factor $\lambda$ on heating a circular system of directors, that is $\alpha = \pi/2$. The in-material radius $r$ extends by $\lambda^{-\nu}$, since it is perpendicular to the director and $\nu$ is the opto-thermal Poisson ratio. The result is a cone, also visualised in (c). (d) Combination of circular director patterns with interfaces will result in combination of cones -- see text.}
	\label{fig:cones}
\end{figure}

Here, we move beyond simple square arrays of cones, and stitch together liquid crystal (LC) defects in more complex patterns, leading to flat LC solid sheets that actuate into sophisticated surfaces. Our basic building blocks are simple $m = +1$ defect patterns consisting of  concentric circles, as shown in Fig.~\ref{fig:cones}.  In isolation, these patterns indeed actuate to form cones, with the evolving Gaussian curvature (GC) localised at the cone tip which is at the defect centre.
%\begin{figure}[!th]
%	\centering
%	\includegraphics[width=\columnwidth]{jpg/cones.jpg}
%	\caption{(a) A director field $\bfn(\bfr)$ making an angle $\alpha$ to the radial vector $\bfr$ itself at angle $\theta$. (b) A circumference $2\pi r$ contacts by a factor $\lambda$ on heating a circular system of directors, that is $\alpha = \pi/2$. The in-material radius $r$ extends by $\lambda^{-\nu}$, since it is perpendicular to the director and $\nu$ is the opto-thermal Poisson ratio. The result is a cone, also visualised in (c). (d) Combination of circular director patterns with interfaces will result in combination of cones -- see text.}
%	\label{fig:cones}
%\end{figure}
This paper is concerned therefore with defects, but since the regions of concentric circles  must meet each other, we are also necessarily concerned with interfaces [Fig. \ref{fig:cones}(d)]. In general these curves of meeting are curved in the still-flat reference state and, on actuation, carry GC varying along their length. Where in turn these interfacial lines meet, there are points where the director field has  features similar to those of negative topological charges $m = -1/2, -1, -3/2, \dots$, depending on how many circular field regions ($3, 4, 5, \dots$) meet at the point. Such points, and their surrounding director fields, are distorted forms of classic LC defects that we analyse in detail elsewhere for their forms and distributions of GC.

The landscape, on actuation, is thus Gauss flat except for concentrations of positive GC which form cone-like tips,  lines carrying (generally) negative GC which form curved folds, and localised GC at points where these folds meet. The typical result is a surface where curved folds traverse saddles between positively curved tips.

Conventional isometric origami involves forming a two-dimensional (2-D) sheet into a three-dimensional (3-D) surface by imposing fold lines. Such origami is mechanically limited since the working material (paper) is unstretching, so the resultant surfaces cannot be Gauss curved. By analogy, non-isometric origami \cite{modes2011blueprinting,MWSPIE:12, Paul_non-isometric} is the evolution of structures via folding at boundaries between director fields, but when length changes of the medium are the driver. Unlike isometric origami, this enables strong, GC bearing surfaces.  In conventional origami, the introduction of curved folds greatly enriches the diversity of possible structures, and can lend them considerable strength since their deformations become highly constrained in order to avoid developing GC \cite{fuchs1999more,dias2012geometric,Choma_2019}.  However, previous work on non-isometric origami considers patterns of piecewise-constant director patches, which are required, by metric compatibility, to meet at straight lines. These lines become straight folds in the actuated state, with the only GC concentrated at points where such lines intersect. Here, we  instead build from patches of concentric circles, and find that metric compatibility between regions leads to curved interfaces and  non-isometric curved-fold  origami. Unlike isometric curve-fold origami, the resultant creases are Gaussian curved and are intrinsic to the geometry of the final surface.

In this paper, we first introduce director fields that actuate to cones, and show how to stitch them to each other in a mechanically compatible way. The resultant boundaries transpire to be hyperbolae and ellipses in the flat sheet, giving rise to two elementary types of double-cone patterns. Hyperbolic interfaces can be combined to produce quite complex groups of cones, which then serve a building blocks for regular and irregular tilings of the plane.  Interestingly, the number of degrees of freedom are  different in tilings of different topologies: some tilings are completely specified by their boundaries and some have freedoms in their interior. In the final part of the paper, we exploit the latter freedom to create differing types of activated topographies of cone tips, and demonstrate how to design a flat sheet that morphs into a desired ``pixelated" surface of cone tips.

In this work we focus on exact analytic embeddings of the activated surfaces, derived as combinations of cones. However, it is likely that other isometries exist, with bend energies acting as a tie-break between them. Bend will also soften and diffuse some of the sharp actuated features we describe. In future work we will return to such bend problems using finite element analysis in the context of a wider analysis of possible structures.  We also reserve for future work a direct calculation of the concentrated Gauss curvature encoded in  curved folds, which, we anticipate,  will be greatly facilitated by recent work describing the geometry of director fields in terms of their splay and bend \cite{niv2018geometric}, and subsequent deployments of this technology to calculate  distributed \cite{griniasty2019curved} and concentrated  \cite{Biggins_curvature_2020} Gauss curvature's on activation. Within the scope of the current work, access to analytic embeddings renders such calculations are largely immaterial, but they are likely to be vital to understanding the forms of curved folds encoded in more general nematic patterns.

\section{Deformations and point sources of Gaussian curvature}

Consider a director field $\bfn(x,y) = n_1(x,y)\, \bfe_1 + n_2(x,y)\, \bfe_2$ at position $(x, y) \in \calD \subset \mathbb{R}^2$ satisfying $|\bfn(x,y)|=1$, where $\{\bfe_1, \bfe_2\}$ are the standard orthonormal Cartesian coordinates in $\mathbb{R}^2$. Upon stimulation, there is a spontaneous deformation with local deformation gradient
\beq
\bfU_{\bfn} = \lambda \bfn \otimes \bfn + \lambda^{-\nu} \bfn^{\perp} \otimes \bfn^{\perp}, \label{eq:gradient}
\eeq
where $\bfn$ is the director and $\bfn^{\perp} = -(\bfn\cdot\bfe_2) \bfe_1 + (\bfn\cdot\bfe_1) \bfe_2$ is the unit  perpendicular to the director.
Under $\bfU_{\bfn}$, the elastomer sheet has a contraction $\lambda<1$ along $\bfn$ and an elongation $\lambda^{-\nu}>1$ along $\bfn^{\perp}$ with the optothermal Poisson ratio $\nu$. Despite having the intrinsic metric induced by $\bfU_{\bfn}$, it is still hard to determine the deformed shape for general director patterns because of the lack of bending information. With the help of symmetry, in particular circular symmetry, deformed shapes such as cones, spherical caps, and more general surfaces of revolution \cite{warner2018nematic}, along with their director patterns, have been described  \cite{modes2010disclinations, aharoni2014geometry, Mostajeran2015,mostajeran2016encoding}.

{\it Point source of Gaussian curvature.} We are interested in point sources of Gaussian curvature, which are specifically cones in the deformed domain.
With the cone angle $\varphi$, the integrated Gaussian curvature is $2\pi(1-\sin\varphi)$. It is localised at the tip, and zero elsewhere on the surface, and represents the angular deficit at this vertex.
%The cone-like structures formed by actuating a nematic elastomer sheet with circular or spiral director patterns have been exhaustively studied in \cite{warner2018nematic, mostajeran2016encoding, mostajeran2017frame}
% An efficient heavy load lifter \cite{white2015programmable}.
We restrict ourselves to director circles since the union of the resulting target space cones can be treated analytically. Logarithmic spiral patterns also actuate to cones, but there are accompanying shears and rotations \cite{mostajeran2017frame}. The rotation itself has significant effects on the deformed domain, especially on the deformed interfaces. We show in another paper that the mechanically compatible director field unions can be analytically calculated, but not various  associated isometries.

The director field $\bfn(r, \theta)$ in polar coordinates, Fig.~\ref{fig:cones}(c), is
\beq
\bfn(r,\theta) =\pm \bfe_{\theta},
\eeq
where $\bfe_{\theta}=-\sin\theta\, \bfe_1 + \cos\theta\, \bfe_2$.

{\it Cone deformation.}
From Fig.~\ref{fig:cones}(b) one sees that the contraction of the circumference $2\pi r$ by $\lambda$ and the increase of the in-material radius to $\lambda^{-\nu} r$ are only
geometrically consistent with the symmetry and with vanishing GC if a cone in $\mathbb{R}^3$ forms after actuation. Then the deformation describing the actuation from reference domain to deformed domain, which we call the {\it cone deformation}, is a map from $\mathbb{R}^2$ to $\mathbb{R}^3$ (with Cartesian coordinate  $\{\hat{\bfe}_1,\hat{\bfe}_2, \hat{\bfe}_3 \}$). From the right triangle in  Fig.~\ref{fig:cones}(b), it is clear that the cone angle is given by $\sin\varphi = \lambda^{1+\nu}$, giving a simple expression for the cone height. Equally, the in-space radius must be $\lambda r$ in order to generate the new circumference. This is all encoded in the cone deformation:
\beq
\bfy_{c}(\bfr):=\bar{\bfy}_c(r,\theta)=\lambda r ( \hat{\bfe}_r -\cot\varphi\ \hat{\bfe}_3),\label{eq:deformation2}
\eeq
where $\hat{\bfe}_r = \cos\theta\, \hat{\bfe}_1 + \sin\theta\, \hat{\bfe}_2$.

%\vspace{-.4cm}
\section{Metric-compatible interfaces between two circular patterns}
%\vspace{-.4cm}
\subsection{Metric compatibility}
\begin{figure}[h]
	\centering
	\includegraphics[width=\columnwidth]{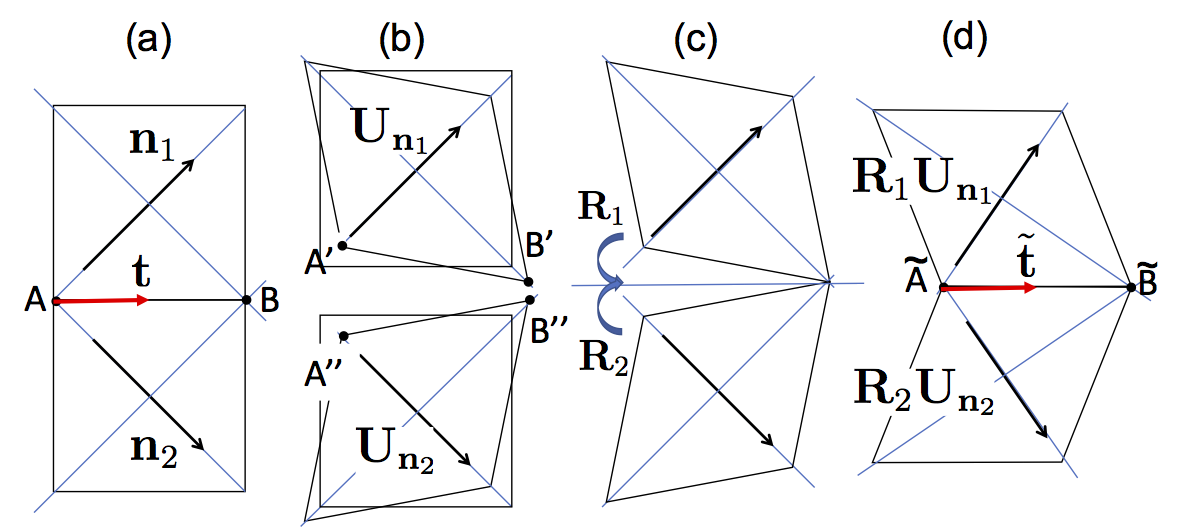}
	\caption{Compatibility between two constant director patterns. Two constant director patterns (a) can be deformed by (b) the stretch tensors $\bfU_{\bfn_i}$ and (c) follow-up rotations to achieve (d) continuous deformed states. $\bft$ and $\tilde{\bft}$ are the tangents to reference and deformed interfaces respectively. }
	\label{fig:rank1}
\end{figure}

Compatibility in continuum mechanics concerns the continuity of a solid body after non-uniform deformations, even when the deformation gradients are discontinuous across an interface in the body. Such continuity, {\it Rank-1 (R-1) compatibility}, or {\it Hadamard's compatibility}, is assured if the deformation gradients at the interface are rank-1 connected, that is satisfy:
\beq
\bfR_1 \bfU_{\bfn_1} - \bfR_2 \bfU_{\bfn_2}  = \bfa \otimes \bft^{\perp}, \label{eq:rank1_2}
\eeq
where $\bfa, \bft, \bft^{\perp} \in \mathbb{R}^2$,  $\bft^{\perp} = -(\bft\cdot\bfe_2) \bfe_1 + (\bft\cdot\bfe_1) \bfe_2$, and $\bft$ is the unit tangent to the interface.
The machinery is  widely used to study martensitic phase transformations \cite{bhattacharya2003microstructure, song2013enhanced}. For deformations of LCE films, the same ideas have been employed \cite{Verwey_96,Finkelmann_97,modes2011blueprinting,MWSPIE:12}. The two patterns in Fig. \ref{fig:rank1} have constant director fields $\bfn_1$ and $\bfn_2$ respectively.
%\begin{figure}[h]
%	\centering
%	\includegraphics[width=\columnwidth]{jpg/rank1.jpg}
%	\caption{Compatibility between two constant director patterns. Two constant director patterns (a) can be deformed by (b) the stretch tensors $\bfU_{\bfn_i}$ and (c) follow-up rotations to achieve (d) continuous deformed states. $\bft$ and $\tilde{\bft}$ are the tangents to reference and deformed interfaces respectively. }
%	\label{fig:rank1}
%\end{figure}
By applying the stretch tensor $\bfU_i$ defined by Eq.~(\ref{eq:gradient}), the two patterns will deform accordingly and a gap between them will emerge. To achieve continuity and merge the gap,
two proper rotations $\bfR_1$ and $\bfR_2$ are needed. The condition Eq.~(\ref{eq:rank1_2}) can be dotted with $\bft$ to give an equivalent condition on the deformation gradients, $\bfR_1 \bfU_{\bfn_1}$ and $\bfR_2 \bfU_{\bfn_2}$:
\beq
(\bfR_1 \bfU_{\bfn_1} - \bfR_2 \bfU_{\bfn_2} ) \bft =0. \label{eq:rank1}
\eeq
A (necessary)  {\it metric compatibility} condition for continuity arises that is useful for studying LCEs since the rotations $\bfR_i$ are usually unable to be uniquely determined locally when the director pattern is non-uniform. Taking the metric $\bfU_{\bfn_i}^T \bfU_{\bfn_i}$ induced by $\bfU_i$, we can eliminate the rotation terms to give compatibility in terms of the metric
\beq
\bft \cdot \bfU_{\bfn_1}^T \bfU_{\bfn_1} \bft  = \bft \cdot \bfU_{\bfn_2}^T \bfU_{\bfn_2} \bft \Leftrightarrow (\bfn_1\cdot \bft)^2 =(\bfn_2 \cdot \bft)^2, \label{eq:metric}
\eeq
which is the condition for the existence of such unknown rotations.

The metric condition (\ref{eq:metric}) implies that locally the lengths of deformed interfaces from two different sides are identical. Also notice that the solutions to (\ref{eq:metric}) are always in pairs, say, $\bfn_1 \cdot \bft = \pm \bfn_2 \cdot \bft$. Then designing a pattern with piecewise constant director fields separated by metric-compatible straight lines in non-isometric origami can be reduced to a limited number of simple rules \cite{modes2011blueprinting,MWSPIE:12, Paul_non-isometric}.

Moreover, we can employ the generalized versions of (\ref{eq:rank1}) and (\ref{eq:metric}) to study curved interfaces, with $\bft$ not a constant vector, but a local tangent to the reference curved interface. This generalization has been used to study the phase transformation and compatible interfaces between helical structures \cite{feng2019phase}. Specifically, suppose the reference interface is described as $\bfr(s)$. Then the generalized version of (\ref{eq:metric}) is
\beq
(\bfn_1(s)\cdot \bft(s))^2 = (\bfn_2(s) \cdot \bft(s))^2, \label{eq:metric_gen}
\eeq
where $\bft(s) = \bfr'(s)/|\bfr'(s)|$. The generalized metric compatibility is employed to study curved interfaces between circular patterns below.

\subsection{Circle/circle interface} \label{subsec:circlecircle}
\begin{figure}[!th]
	\centering
	\includegraphics[width=\columnwidth]{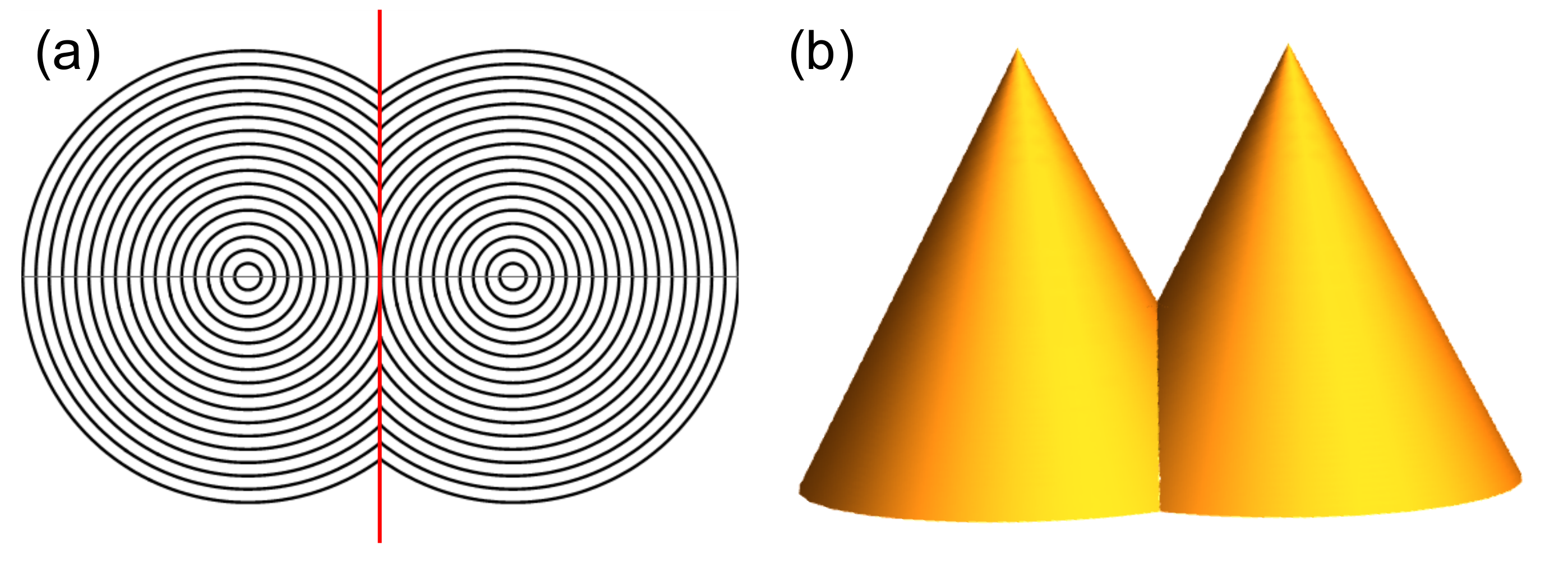}  %%small in any event, no need to provide small version
	\caption{An example of straight-line interface (red) between two symmetrically located circular patterns. (a) Two circular patterns are separated by a straight-line interface (red). (b) The deformed configuration has positive GC at the tips and negative GC along the deformed interface. }
	\label{fig:symmetric_circle}
\end{figure}

Two circular director patterns can have metric-compatible interfaces between them. The shape of the interface depends on the branch of solution to Eq. (\ref{eq:metric}) and on where the interface passes through the line connecting the pattern centres.  As a starting point, Fig. \ref{fig:symmetric_circle} provides an example of a straight-line interface between two circular patterns that bisects the line connecting the two centres \cite{MWSPIE:12}.
%\begin{figure}[!th]
%	\centering
%	\includegraphics[width=\columnwidth]{jpg/symmetric_circle.jpg}
%	\caption{An example of straight-line interface (red) between two symmetrically located circular patterns. (a) Two circular patterns are separated by a straight-line interface (red). (b) The deformed configuration has positive GC at the tips and negative GC along the deformed interface. }
%	\label{fig:symmetric_circle}
%\end{figure}
The deformed domain [Fig. \ref{fig:symmetric_circle}(b)] has two equal-height tips with concentrated positive GC and a curved interface with distributed negative GC. The straight-line interfaces between circular patterns have been widely used to design arrays of LCE lifters and complex topography, even with multi-layers \cite{mcconney2013topography, ware2015voxelated}.

One can estimate the integrated GC in the curved interface: at large distances the twin cone has a field close to that of a single cone. The angular deficit apparent from the far field must thus be $2\pi(1 - \sin\varphi)$. Thus integrated GC equivalent to that in one tip must have been cancelled by that residing in the curved interface.

\begin{figure}[!th]
	\centering
	\includegraphics[width=\columnwidth]{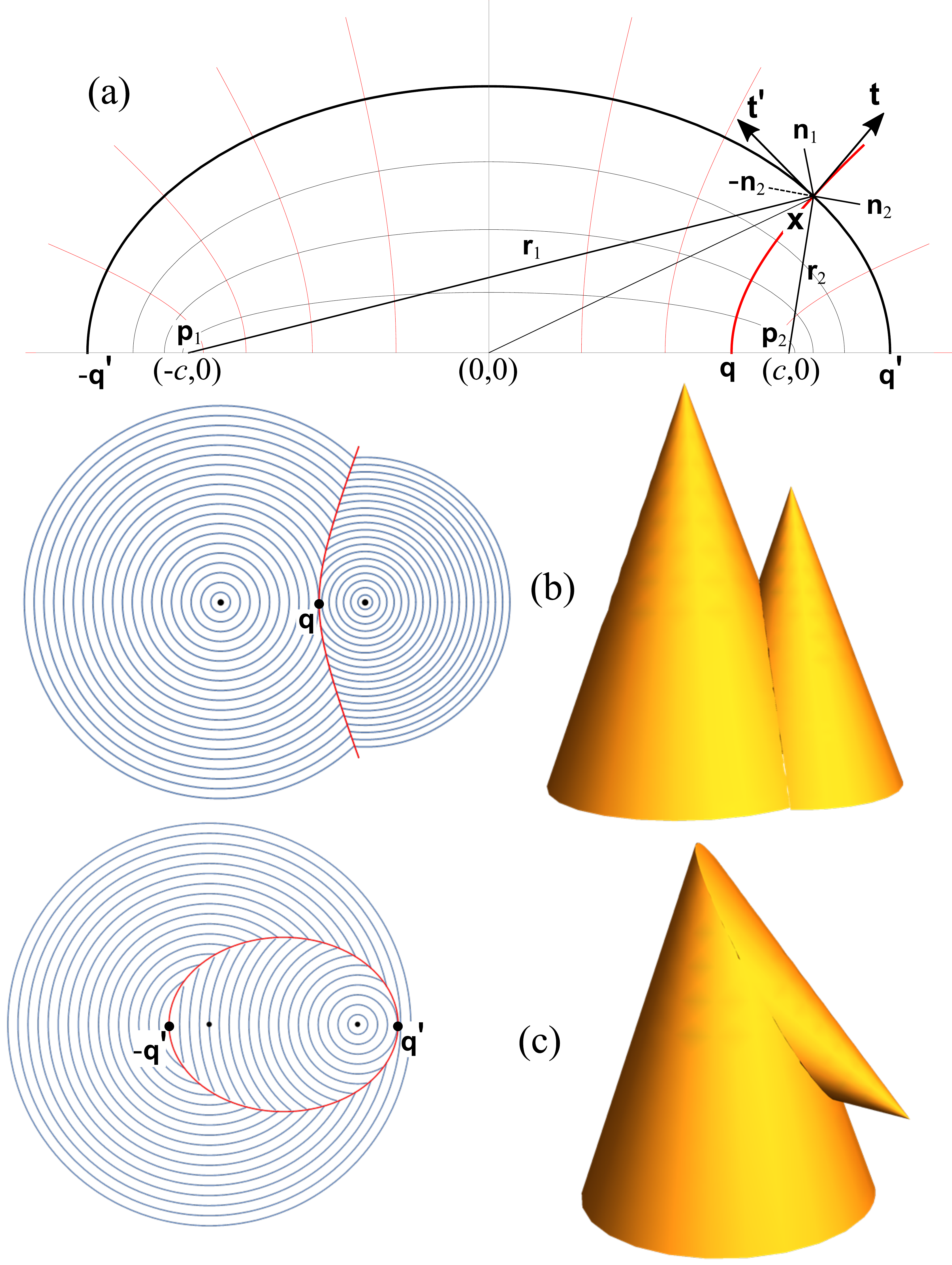}
	\caption{Two types of circle/circle interfaces. (a) The centres of the circle systems, $\bfp_i$ are the foci of the system of ellipses and hyperbolae that form the two families of possible R-1 connected interfaces.  The offset $\bfq$ is the intersection between a particular hyperbolic interface and the horizontal axis. At the point $\bfx$, the hyperbola, with tangent $\bft$ is seen to bisect the directors $\bfn_1$ and $\bfn_2$ based on systems 1 and 2 of circles, while ellipse with tangent $\bft'$ also passing through $\bfx$ is seen to bisect $\bfn_1$ and $-\bfn_2$. (b) Hyperbolic interface: The deformed domain (right) is two connected partial cones with the same axis but different heights, deformed from the reference domain (left) by two cone deformations. (c)  Elliptical interface: The deformed domain (right) is two connected partial cones with different axes and heights, deformed from the reference domain (left).}
	%\caption{Two types of circle/circle interfaces. (a) The coordinates of the system. The offset $\bfq$ is the intersection between the interface and the horizontal axis. (b) Hyperbolic interface: The deformed domain (right) is two connected partial cones with the same axis but different heights, deformed from the reference domain (left) by two cone deformations. (c) The family of hyperbolic interface.
	%		(d) Elliptical interface: The deformed domain (right) is two connected partial cones with different axes and heights, deformed from the reference domain (right). (e) The family of elliptical interface.}
	\label{fig:circlecircle}
\end{figure}

Generally, the circle/circle interface is not limited to straight lines -- its shape depends on the offset $\bfq$ from the bisection, see Fig. \ref{fig:circlecircle}(a).
It is an elementary property of hyperbolae that at a point on it, its tangent vector bisects the angle made between the radial vectors $\bfr_i$ from the two foci to this point $\bfx$, here from the two director pattern centres located at $\bfp_1= (-c, 0)$ and $\bfp_2=(c, 0)$. Since the directors associated with $\bfr_1$ and $\bfr_2$ are both at right angles to the radii, the angle between the directors will also be bisected. The equivalent property of ellipses is that the tangent to the ellipse passing through a given point makes equal angle to the radial vectors. It is thus the bisector of the complementary angle between the two directors, that is it represents the other solution to Eq.~(\ref{eq:metric_gen}).

%A striking property about circle/circle interface is that the shapes of deformed metric compatible interfaces, formed by cone deformations associated with the two sources,  are essentially the same\footnote{We can match one deformed interface to another by applying some constant rotation and translation.}. Then the deformed shapes from both sides remain on cones. In this section, we study the reference  and deformed interfaces between two circular patterns.
To calculate the metric-compatible interface analytically and exhaustively, we establish the coordinates in Fig. \ref{fig:circlecircle}(a) with the directors $\bfn_1$ and $\bfn_2$ at $\bfx$ from two sides of the interface
\beq
\bfn_1 = \bfR(\pi/2) \frac{\bfr_1}{|\bfr_1|}, ~ \bfn_2 = \bfR(-\pi/2) \frac{\bfr_2}{|\bfr_2|},
\eeq
where $\bfR(.)=\cos(.) (\bfe_1 \otimes \bfe_1 + \bfe_2 \otimes \bfe_2) + \sin(.) (-\bfe_1 \otimes \bfe_2 + \bfe_2 \otimes \bfe_1)$ is a rotation in $\mathbb{R}^2$ and $\bfr_i = \bfx - \bfp_i, i=1,2$. The tangent $\bft$ of the interface, as depicted in Fig. \ref{fig:circlecircle}(a), is the bisector between $\bfn_1$ and $\bfn_2$, and satisfies $\bfn_1 \cdot \bft = \bfn_2 \cdot \bft$. The other tangent $\bft'$ is another solution corresponding to $\bfn_1 \cdot \bft =- \bfn_2 \cdot \bft$ and perpendicular to $\bft$. Here we should notice that $\bfn_1$, $\bfn_2$, and $\bft$ are not constants. To solve for the interface, we find it convenient to write $\bfx$ in elliptic coordinates as $\bfx= (c \cosh u \cos v, c \sinh u \sin v)$. The facts that $\bft$ bisects $\angle \bfp_1 \bfx \bfp_2$ and $\bft' \perp \bft$ yield two families of solutions:
\begin{enumerate}
	\item Hyperbolic interface [Fig. \ref{fig:circlecircle}(a) \& (b)]. The parametric form of the interface is
	\beq
	\begin{aligned}
		\begin{cases}
			x(u)=c \cosh u \cos v_0 \\
			y(u)=c \sinh u \sin v_0
		\end{cases}
	\end{aligned},  \label{eq:hyper}
	\eeq
	where $v_0 \in (0, \pi)$ is a constant and $u \in \mathbb{R}$ varies parametrically along the hyperbola.
	The offset $\bfq$ is at $\bfq=(c \cos v_0, 0)$.
	\item Elliptical interface [Fig. \ref{fig:circlecircle}(a) \& (c)]. The parametric form of the interface is
	\beq
	\begin{aligned}
		\begin{cases}
			x(v) = c \cosh u_0 \cos v \\
			y(v)=c \sinh u_0 \sin v
		\end{cases},
	\end{aligned}
	\eeq
	where $u_0 \in \mathbb{R}\setminus {0}$ is a constant and $v \in [0, 2\pi)$ is the parametric parameter. The offset $\bfq'$ is at $\bfq'=(c \cosh u_0, 0)$.
\end{enumerate}
By varying the offsets $\bfq$ and $\bfq'$, we obtain two families of interfaces depicted in Fig. \ref{fig:circlecircle}(a). The elliptical division of the plane is qualitatively different since the one pattern subsumes the centre of the other, and there is only one cone tip in the activated state.

Next, we calculate the deformed configuration. For the hyperbolic interface, the interfaces deformed from two different sources are
\beqs
\bfy_1(u)&=&\bfy_{c}(\bfr_1(u)), \\ \nonumber
\bfy_2(u)&=&\bfy_{c}(\bfr_2(u))
\eeqs
respectively, where
\beqs
\bfr_1(u)&=&c (\cosh u \cos v_0 + 1, \sinh u \sin v_0), \\  \nonumber
\bfr_2(u)&=&c (\cosh u \cos v_0 - 1,  \sinh u \sin v_0).
\eeqs
The moduli $|\bfr_1(u)|$ and $|\bfr_2(u)|$ are $\cosh u \pm \cos v_0$, whence Eq.~(\ref{eq:deformation2}) and differentiation gives the tangent of the deformed interface.  For instance:
\beq
 \bfy'_1(u) = \lambda c (\cos v_0 \sinh u, \sin v_0 \cosh u, - \cot \varphi \sinh u).
\eeq
The tangent  lies on the plane perpendicular to $(\cos\varphi, 0, \sin\varphi \cos v_0)$. Then the deformed interface is a hyperbola, since it is a planar section of a cone with the angle between the cone axis and the plane less than the cone angle $\varphi$.
Furthermore, by direct calculations, we have
\beq
\bfy_1(u) - \bfy_2(u) = 2 \lambda c (\hat{\bfe}_1 - \cos v_0 \cot \varphi \hat{\bfe}_3), \label{eq:trans1}
\eeq
which is a constant translation, independent of $u$, which allows the two cones separately evolving from their own circular patterns to be bodily translated to assure the join between them in the target space. That is, no isometry additional to this translation is required. The $\hat{\bfe}_1$ translation is simply seen from Fig.~\ref{fig:cones}(b) where the in-space distances $\lambda r$ are contracted by $\lambda$ from their reference state values. Here the contraction is of the initial distance $2 c$  connecting the centres.  See also the $\hat{\bfe}_r$ term of Eq.~(\ref {eq:deformation2}).
The relative height, which is the height of $\bfp_1$ minus the height of $\bfp_2$ on the deformed domain, is
\beqs
\Delta h &=& -(\bfy_1(0) - \bfy_2(0)) \cdot \hat{\bfe}_3 = 2 \lambda c  \cos v_0  \cot \varphi\label{eq:height_a}\\
&=& \lambda^{-\nu}  (|\bfp_1 -\bfq| - |\bfp_2 -\bfq|) \cot \varphi. \label{eq:height}
\eeqs
Heights are seen from the right triangle of Fig.~\ref{fig:cones}(b) to be $\lambda \cot\varphi$ times the initial in-plane distance ($r$ in that figure); see also Eq.~(\ref {eq:deformation2}). The height differences, relative to the saddle, depend on the difference of the reference state distances to $\bfq$, that is $c(1+\cos v_0) - c(1-\cos v_o) = 2c\cos v_0$, which is what enters Eq.~(\ref{eq:height_a}) and is compactly expressed in Eq.~(\ref{eq:height}). Note $ \lambda \cot \varphi  \equiv  \lambda^{-\nu} \cos\varphi $.
Thus the relative height depends on $\bfp_1$, $\bfp_2$, and $\bfq$, which is later useful to explore inverse design principles.
The remarks made for the bisected case about the integrated GC residing in the curved interface between the cones apply equally here.

For the elliptical interface, the deformed interfaces from two sources are
\beqs
\bfy_1(v)&=&\bfy_{c}(\bfr_1(v)), \\ \nonumber
\bfy_2(v)&=&\bfy_{c}(\bfr_2(v))
\eeqs
respectively, where
\beqs
\bfr_1(v)&=&c (\cosh u_0 \cos v + 1, \sinh u_0 \sin v), \\ \nonumber
\bfr_2(v)&=&c (\cosh u_0 \cos v - 1,  \sinh u_0 \sin v).
\eeqs
The tangent of the deformed interface,
	\beq
	 \bfy'_1(v) = \lambda c (-\cosh u_0 \sin v, \sinh u_0 \cos v,  \cot \varphi \sin v),
	\eeq
	lies on the plane perpendicular to $(\cos\varphi, 0, \sin\varphi \cosh u_0)$. Then the deformed interface is an ellipse, since it is a planar section of a cone with the angle between the cone axis and the plane greater than the cone angle $\varphi$.

Let the axis of the cone evolving from the left pattern remain parallel to $\hat{\bfe}_3$. The deformed interface $\bfy_1(v)$ is an ellipse as an inclined plane cut of the cone. The second interface $\bfy_2(v)$ is also an ellipse, but its orientation is different from $\bfy_1(v)$. Then a rotation is required to match these two deformed interfaces.
By direct calculation, explicitly, the following equality holds:
\beq
\bfy_1(v) - \bfR_{\hat{\bfe}_2} (\xi) \bfy_2(v) = \frac{2 \lambda c \cosh^2 u_0 \csc^2 \varphi}{\cosh^2 u_0 + \cot^2 \varphi} (\hat{\bfe}_1-\frac{\cot\varphi}{\cosh u_0} \hat{\bfe}_3), \label{eq:trans2}
\eeq
where $\xi$ is a constant rotation angle uniquely determined by the well-defined functions
\beq
\begin{aligned}
	\begin{cases}
		\sin\xi=\frac{2 \cosh u_0 \cot\varphi}{\cosh^2 u_0 + \cot^2 \varphi} \\
		\cos \xi = \frac{\cosh^2 u_0 - \cot^2 \varphi}{\cosh^2 u_0 + \cot^2 \varphi}
	\end{cases},
\end{aligned}
\eeq
and $\bfR_{\hat{\bfe}_2} (.) := \cos(.)(\hat{\bf e}_1 \otimes \hat{\bf e}_1 + \hat{\bf e}_3 \otimes \hat{\bf e}_3 ) + \sin(.) (\hat{\bf e}_1 \otimes \hat{\bf e}_3 - \hat{\bf e}_3 \otimes \hat{\bf e}_1  ) +\hat{\bf e}_2 \otimes \hat{\bf e}_2 $ is a rotation tensor about $\hat{\bfe}_2$. The $\bfR_{\hat{\bfe}_2} (\xi)$ in (\ref{eq:trans2}) is essentially the rotation needed to reorient the deformed interface $\bfy_2(v)$ in order to match the orientation of $\bfy_1(v)$. The right-hand side of (\ref{eq:trans2}) is also a constant translation. Similar to the hyperbolic case, no isometry additional to the rotation and translation is required to join the interfaces.

The integrated GC in the deformed elliptical interface now vanishes: The director field of the composite object, and indeed its activated shape, is just that of a cone if one is beyond the interface between the two regions. The angular deficit is thus $2\pi(1-\sin\varphi)$. This value is also that associated with the one tip of the actuated object, and hence the GC in the interface must vanish. Inspection of the interface indeed shows regions of positive and negative GC.

To summarize, the circle/circle interfaces in both hyperbolic and elliptical families have the same deformed shapes [Eqs. (\ref{eq:trans1}) and (\ref{eq:trans2})] obtained by the cone deformations corresponding to the two sources. More importantly, the hyperbolic family only needs translations to match the interfaces and the axes of deformed cones are all parallel. This fact inspires us to conveniently tile a surface with circular patterns separated by hyperbolic interfaces in the following sections.

\section{Patterning complex topography from circular director patterns}
%\begin{figure}[]
%	\centering
%	\includegraphics[width=\columnwidth]{jpg/n_fold_long.jpg}
%	\caption{Examples of n-fold intersection: reference domain and deformed configuration. The red curves are the metric-compatible hyperbolic interfaces between circular director patterns. The three-fold intersection (a), four-fold intersection (b), and n-fold intersection (c) have topological charges $-1/2$, $-1$, and  $-(n-2)/2$ respectively.}
%	\label{fig:n_fold}
%\end{figure}
\begin{figure}[]
	\centering
	\includegraphics[width=\columnwidth]{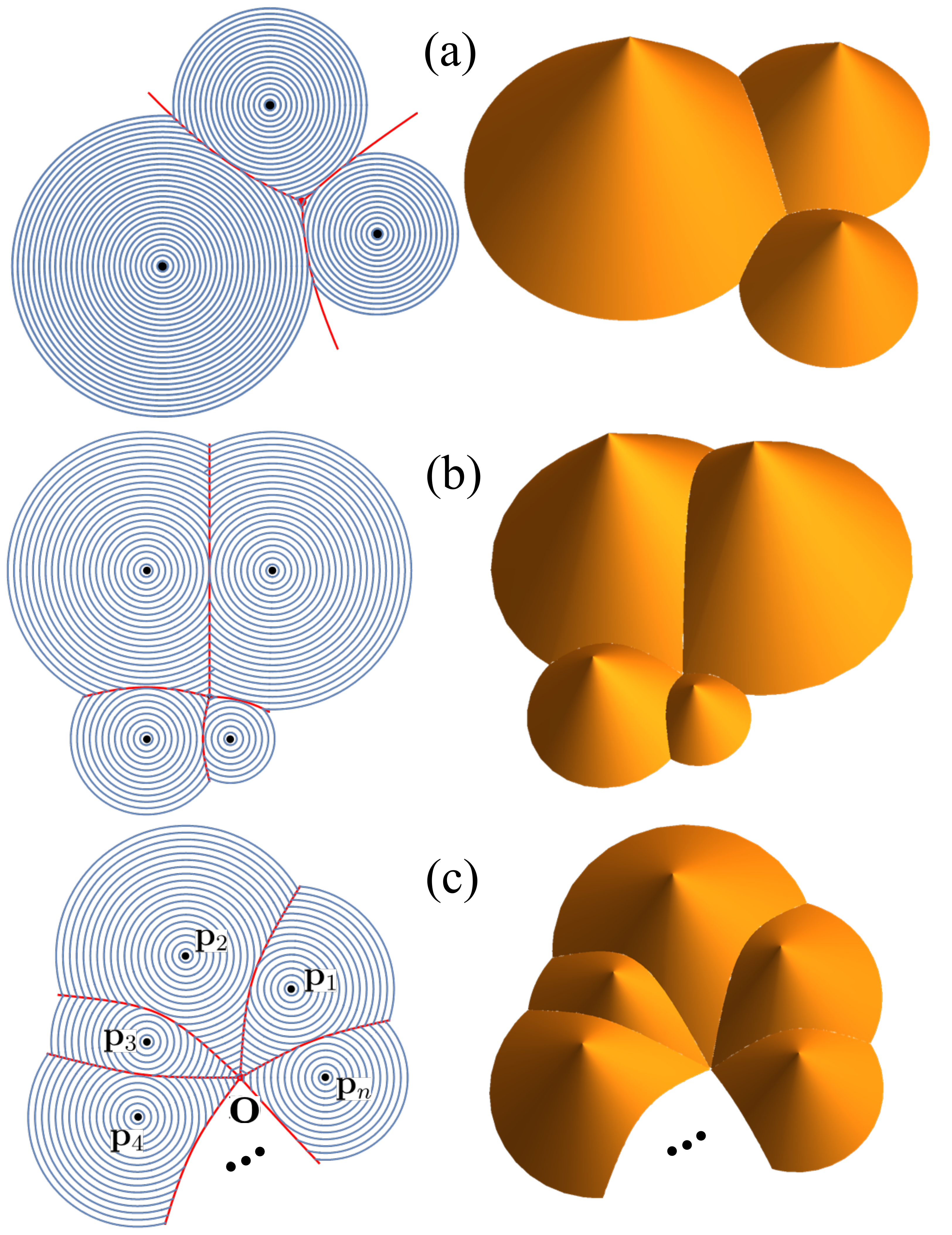}
	\caption{Examples of n-fold intersection: reference domain and deformed configuration. The red curves are the metric-compatible hyperbolic interfaces between circular director patterns. The three-fold intersection (a), four-fold intersection (b), and n-fold intersection (c) have topological charges $-1/2$, $-1$, and  $-(n-2)/2$ respectively.}
	\label{fig:n_fold}
\end{figure}
\vspace{-.4cm}
Patterning large-scale structures from building blocks is a typical technique for designing complex functionalities from small basic units, for example, the designs of metamaterials \cite{schenk2013geometry} and soft robotics \cite{kim2018printing}.
In this section, we employ  circular director patterns with hyperbolic interfaces as building blocks of complex topographies respecting global compatibility. The design fundamentally relies on the local compatibility of interfaces between circular patterns, both in the reference and the deformed domains. The ultimate topographies have non-trivial GC concentrated at tips, interfaces, and intersections of interfaces. More importantly, some types of topographies can be inverse designed based on the freedom to pattern these building blocks.

\subsection{Basic building blocks: three-fold, four-fold, and n-fold intersections of circle systems} \label{sec:nfold}
A given (hyperbolic) interface cuts at right angles the line connecting the centres, e.g. $\bfp_1$, $\bfp_2$ of two neighbouring patterns. Likewise centres  $\bfp_2$, $\bfp_3$ and  $\bfp_3$, $\bfp_4$ . . . will be separated by such interfaces. To tile a plane with systems  $\bfp_1$, $\bfp_2$, . . .  we need to determine how the family of interfaces themselves meet at points.

\begin{figure*}[!th]
	\centering
	\includegraphics[width=\textwidth]{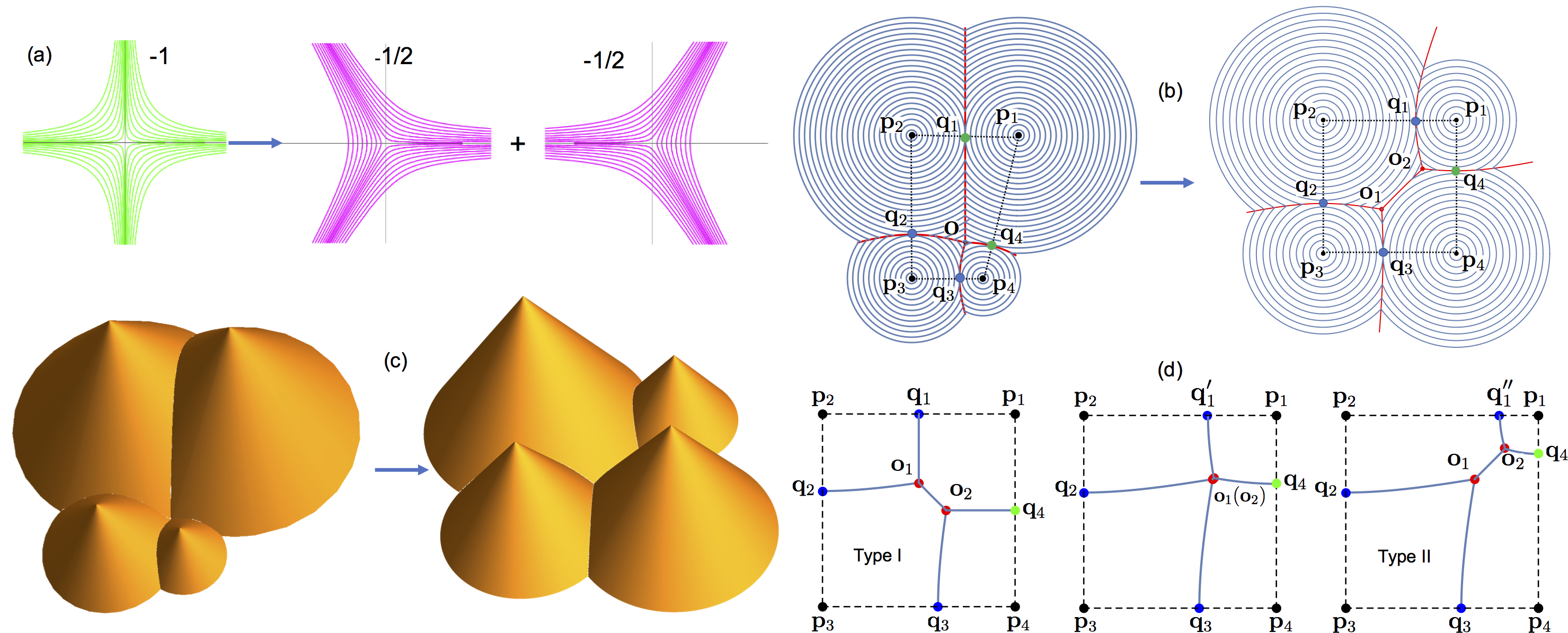}
	\caption{(a) The splitting of topological charge $-1$ into $(-1/2) + (-1/2)$. (b)-(c) The splitting of the four-fold intersection into two three-fold intersections: (b) reference domain and (c) deformed configuration. (d) Type I two connected three-folds (left) passes the critical four-fold state (middle),  and then transforms to Type II (right), as moving $\bfq_1$ closer to $\bfp_1$ but keeping $\bfq_2$, $\bfq_3$ and $\bfp_i$ unchanged.}
	\label{fig:splitting}
\end{figure*}

Given two centres $\bfp_1$ and $\bfp_2$ as the foci located at $(-c, 0)$ and $(c,0)$, there exists a unique hyperbola passing through a prescribed point $ \bfo =(x_0, y_0)\neq (\bar{c}, 0), |\bar{c}|\geq c$.
Specifically, the parameter $v_0$ in (\ref{eq:hyper}) is determined by
\beq
v_0=\arccos \left(  \frac{\sqrt{(x_0+c)^2 + y_0^2} - \sqrt{(x_0 - c)^2 + y_0^2}}{2 c}  \right). \label{eq:v0i}
\eeq
Multiple hyperbolic interfaces separating multiple centres can meet at the same point [$\bfo$ in Fig. \ref{fig:n_fold}(c)] to form three-fold, four-fold, and in general n-fold intersections [Fig. \ref{fig:n_fold}]. These intersections have  features similar to those of
topological charges ($-1/2$, $-1$, and $-(n-2)/2$ respectively) in the nematic director field.
For instance, the $-1$ charge in Fig.~\ref{fig:splitting}(a) has four asymptotic ``folds" similar to the four-fold intersection curves in Fig. \ref{fig:n_fold}(b). Topological charges in liquid crystals are discerned by the winding of director orientation around {\it any} loop containing the defect (see \cite{tang2017orientation} for more details). Our examples in Fig. \ref{fig:n_fold} differ -- for the winding to be well defined, the loop must pass through the interfaces at points of matching director, such as the \textbf{q}$_i$ points of Fig. \ref{fig:splitting}(b)[left]. We explicitly design  n-fold intersections with compatible interfaces in the reference and deformed domains where $n$ centres meet.

{\it Reference domain.} To calculate the analytical forms of the hyperbolic interfaces on the reference domain, we assume the $n$ centres $\bfp_1, \bfp_2, \dots,\bfp_n$ form a convex polygon that contains the intersection $\bfo$, as depicted in Fig. \ref{fig:n_fold}(c). $\bfp_i, \bfo \in \mathbb{R}^2$ denote the position of centres or of the intersection. The interfaces are obtained by rotating the system accordingly, then calculating the interface by (\ref{eq:hyper}) and (\ref{eq:v0i}), and then rotating back. Explicitly, let $\bfR_i$ denote the rotation that rotates $\bfp_{i+1} - \bfp_i$ parallel to $\bfe_1$, i.e.,
$\bfR_i = \bfe_1 \otimes \bft_i + \bfe_2 \otimes \bft_i^{\perp}$, where $\bft_i =\frac{\bfp_{i+1}-\bfp_i}{|\bfp_{i+1}-\bfp_i|}$, $\bft_i^{\perp} = -(\bft_i \cdot\bfe_2) \bfe_1 + (\bft_i \cdot\bfe_1) \bfe_2$, and $\bfp_{n+1} :=\bfp_1$. In order to calculate the interface between $\bfp_i$ and $\bfp_{i+1}$, we rotate and translate the system to have the new foci at $(-c_i, 0)$ and $(c_i, 0)$ with $c_i =  |\bfp_{i+1} - \bfp_i|/2$, and the new intersection at $(x_{0i}, y_{0i})$ with
\begin{align}
\begin{cases}
x_{0i} = \bfR_i (\bfo - \frac{\bfp_i + \bfp_{i+1}}{2}) \cdot \bfe_1 \\
y_{0i} = \bfR_i (\bfo - \frac{\bfp_i + \bfp_{i+1}}{2}) \cdot \bfe_2
\end{cases}.
\end{align}
Then the hyperbolic interface between $\bfp_i$ and $\bfp_{i+1}$ has the parametric form
\beq
\bfh_i(u) = \bfR_i^T (x_i(u), y_i(u))  + \frac{\bfp_{i+1} + \bfp_i}{2},\quad i=1,\dots,n, \label{eq:nfold}
\eeq
where
\begin{align}
	\begin{cases}
		x_i(u) = c_i \cosh u \cos v_{0i} \\
		y_i(u) = c_i \sinh u \sin v_{0i}
	\end{cases},
\end{align}
with the parameter
\beq
v_{0i}=\arccos \left(  \frac{\sqrt{(x_{0i}+c_i)^2 + y_{0i}^2} - \sqrt{(x_{0i} - c_i)^2 + y_{0i}^2}}{2 c_i}  \right)
\eeq
calculated as for (\ref{eq:v0i}).

%\begin{figure*}[!th]
%	\centering
%	\includegraphics[width=\textwidth]{jpg/splitting_2.jpg}
%	\caption{(a) The splitting of topological charge $-1$ into $(-1/2) + (-1/2)$. (b)-(c) The splitting of the four-fold intersection into two three-fold intersections: (b) reference domain and (c) deformed configuration. (d) Type I two connected three-folds (left) passes the critical four-fold state (middle),  and then transforms to Type II (right), as moving $\bfq_1$ closer to $\bfp_1$ but keeping $\bfq_2$, $\bfq_3$ and $\bfp_i$ unchanged.}
%	\label{fig:splitting}
%\end{figure*}

{\it Deformed domain.} According to Sect. \ref{subsec:circlecircle}, specifically Eq. (\ref{eq:trans1}), one can match two deformed hyperbolic interfaces by applying a constant translation.
The translation can be evaluated by matching two deformed points that correspond to the same point on the reference interface. For the n-fold intersection, we match the deformed $\bfo$ for each circular pattern $\Omega_i$ with centre $\bfp_i, i =1,\dots, n$. Then the entire deformed domain is
\beq
\bfy_c(\bfx -\bfp_i) - \bfy_c(\bfo - \bfp_i), ~\bfx \in \Omega_i, ~i=1,\dots,n,
\eeq
where the cone deformation $\bfy_c$ is defined by (\ref{eq:deformation2}).
This setting ensures that the deformed $\bfo$ from different $\bfp_i$ are all at $(0, 0, 0)$, which means all the interfaces are matched perfectly.
Then the deformed structure [Fig. \ref{fig:n_fold}C] is continuous as we expect. In the following sections, we use the similar strategy of applying translations to make the deformed domain continuous.

\subsection{Splitting the order of reference state topological defects} \label{sec:splitting}
We now consider a phenomenon that one four-fold intersection is split into two connected three-fold (CTF) intersections.
For the four-fold intersection [Fig. \ref{fig:splitting}(b) left], the circular pattern centered at $\bfp_2$ has no direct contact with the pattern $\bfp_4$, except for the intersection $\bfo$. But when we increase the sizes of patterns $\bfp_2$ and $\bfp_4$ and retain the positions of offsets $\bfq_2$ and $\bfq_3$, the hyperbolic interface between $\bfp_2$ and $\bfp_4$  emerges, as shown in Figs.~\ref{fig:splitting}(b) \& (c). The $-1$ topological charge associated with the four-fold intersection is then split into two $-1/2$ topological charges associated with the CTFs; the total topological charge is preserved [Fig.~\ref{fig:splitting}(a)]. One can also see that in both figures \ref{fig:splitting}(b), when the circular sectors of the exterior are extended outwards  to large distances, the effective charge is +1 in each case (the outer pattern tending to circular). The contained topological charge has not changed as a result of the fission.
%:
%\beq
%-1=\left(-\frac{1}{2}\right) + \left(-\frac{1}{2}\right).
%\eeq

According to (\ref{eq:nfold}), an n-fold intersection can be uniquely determined by the centres $\bfp_1, \dots, \bfp_n$ and the intersection $\bfo$. The four-fold intersection in Fig. \ref{fig:splitting}(c) follows this method. For the sake of convenience to tile a surface, we introduce here another equivalent approach to construct the four-fold intersection.
We prescribe the centres $\bfp_1, \bfp_2, \bfp_3, \bfp_4$ and the offsets $\bfq_2, \bfq_3$ (blue dots). Recall that the offset $\bfq_i$ is the (perpendicular) intersection between the line connecting two neighboring centres and the hyperbolic interface separating them. Then $(\bfp_2, \bfp_3, \bfq_2)$ and $(\bfp_3, \bfp_4, \bfq_3)$ will determine the hyperbolic interfaces $\widehat{\bfq_2 \bfo}$ and $\widehat{\bfq_3 \bfo}$, and thus the intersection $\bfo$ itself. Once $\bfo$ is determined, the other offsets $\bfq_1, \bfq_4$ (green dots) and the hyperbolic interfaces $\widehat{\bfq_1 \bfo}, \widehat{\bfq_4 \bfo}$ can be calculated. Thus the entire pattern including four centres and four hyperbolic interfaces are determined by the four centres and two offsets. The number of degrees of freedom (DOF) is two, both for the current method
($\bfq_2$ and $\bfq_3$ as offsets on the lines connecting centres), and the previous method ($\bfo$ moving in 2-D), when the centres are given.

For the case of two connected three-fold intersections [Fig. \ref{fig:splitting}(b) right], four centres $(\bfp_1, \bfp_2, \bfp_3, \bfp_4)$ and three offsets $(\bfq_1, \bfq_2, \bfq_3)$  will be needed to determine the entire pattern. Specifically, $(\bfp_2, \bfp_3, \bfq_2)$ and $(\bfp_3, \bfp_4, \bfq_3)$ will determine the hyperbolic interfaces $\widehat{\bfq_2 \bfo_1}$, $\widehat{\bfq_3 \bfo_1}$, and the first intersection $\bfo_1$, the same as the four-fold case. Then, the diagonal hyperbola $\widehat{\bfo_1 \bfo_2}$ determined by $(\bfp_2, \bfp_4, \bfo_1)$ intersects the top hyperbola $\widehat{\bfq_1 \bfo_2}$ determined by $(\bfp_1, \bfp_2, \bfq_1)$ at the second intersection $\bfo_2$. Finally, the fourth hyperbola $\widehat{\bfo_1 \bfq_4}$ and offset $\bfq_4$ is determined by $(\bfp_1, \bfp_4, \bfo_2)$.
In comparison with the four-fold case, this case has one more degree of freedom $-$ the position of offset $\bfq_1$. Recall that the relative height between the deformed $\bfp_1$ and $\bfp_2$ is determined by $(\bfp_1, \bfp_2, \bfq_1)$ [Eq. (\ref{eq:height})]. Then given the four centres $\bfp_i$ and two offsets $\bfq_2, \bfq_3$, the extra DOF of $\bfq_1$ allows us to manipulate the relative height between the deformed $\bfp_1$ and $\bfp_2$, as the basis of the following inverse design.
Technically, determined by the position of $\bfq_1$, there are two types of two connected three-folds [see Fig. \ref{fig:splitting}(d)]:
\begin{itemize}
\item Type I. The intersection $\bfo_1$ is the intersection between the hyperbolae determined by $(\bfp_1, \bfp_2, \bfq_1)$ and $(\bfp_2, \bfp_3, \bfq_2)$. $\bfo_2$ is the intersection between the hyperbolae determined by $(\bfp_1, \bfp_3, \bfo_1)$ and $(\bfp_3, \bfp_4, \bfq_3)$. The hyperbola $\widehat{\bfo_1 \bfo_2}$ is the interface between centres $\bfp_1$ and $\bfp_3$.
\item Type II. The intersection $\bfo_1$ is the intersection between the hyperbolae determined by $(\bfp_2, \bfp_3, \bfq_2)$ and $(\bfp_3, \bfp_4, \bfq_3)$. $\bfo_2$ is the intersection between the hyperbolae determined by $(\bfp_2, \bfp_4, \bfo_1)$ and $(\bfp_1, \bfp_2, \bfq_1^{\prime\prime})$. The hyperbola $\widehat{\bfo_1 \bfo_2}$ is the interface between centres $\bfp_2$ and $\bfp_4$.
\end{itemize}
With the offset $\bfq_1$ moving closer to $\bfp_1$, the pattern undergoes Type I $\rightarrow$ four-fold $\rightarrow$ Type II states successively. The relative height (the height of deformed $\bfp_1$ minus the height of deformed $\bfp_2$) keeps decreasing during the process. The transition between two types of connected 3-folds reminds us the phenomena of topological transition of interfaces in 2-D liquid dry foams, driven by the energy of curved interfaces that follows Plateau's rules \cite{mughal2018demonstration}.

\subsection{Complex topographies with three-fold and four-fold intersections}
\subsubsection{Topographies with two dimensional translation symmetry: objective non-isometric origami}
%\begin{figure}[]
%	\centering
%	\includegraphics[width=\columnwidth]{jpg/symmetry_small.jpg}
%	\caption{Examples of topographies with 2D translation symmetry. The unit cells of centres (rightmost) for the reference and deformed states are (a) a square,  (b) a rhombus, and (c)  a hexagon. Accordingly, the reference (left) and deformed (middle) tilings have (a) four-fold, (b) three-fold, and (c) six-fold intersections of their interfaces. }
%	\label{fig:symmetry}
%\end{figure}
\begin{figure}[]
	\centering
	\includegraphics[width=\columnwidth]{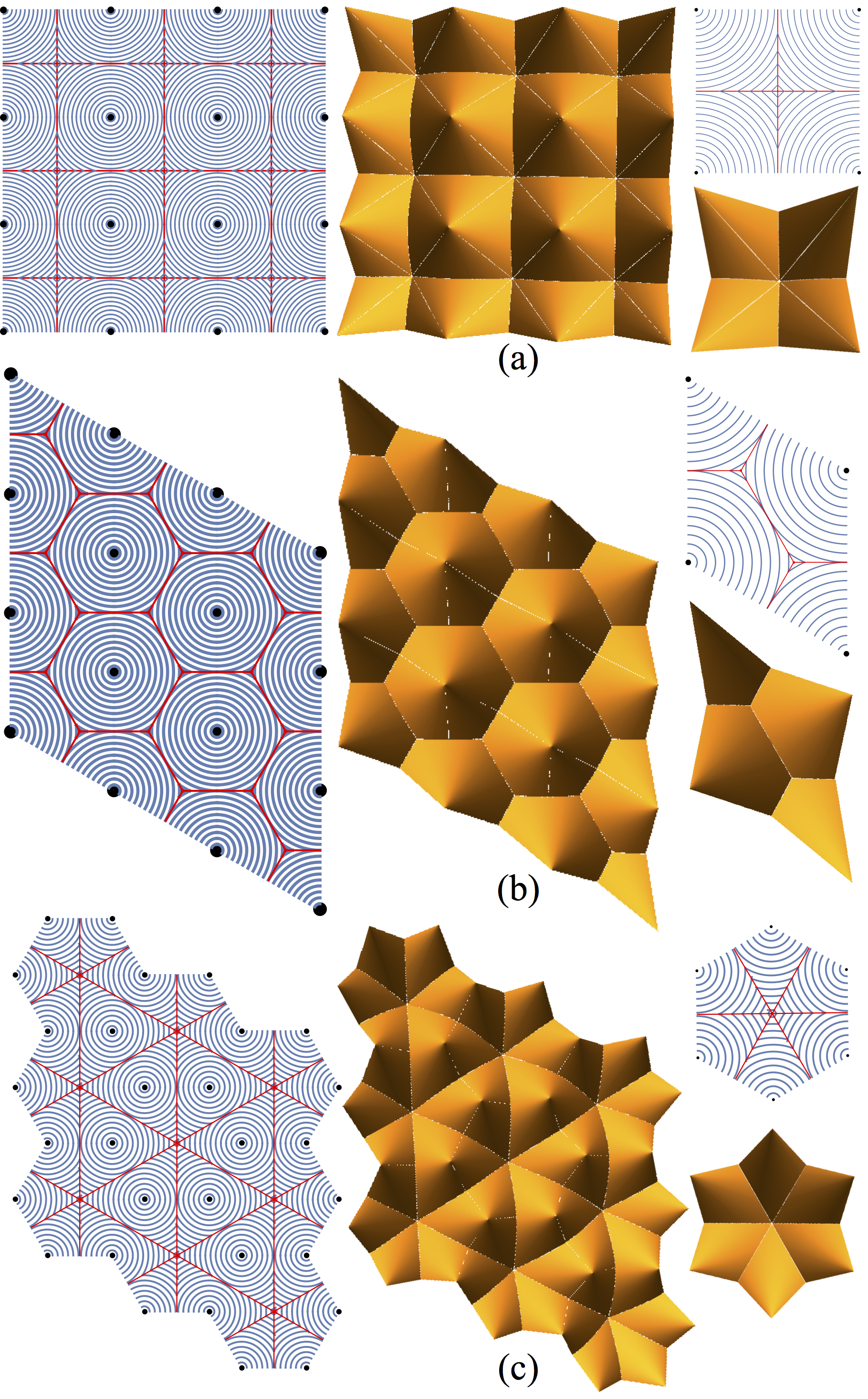}
	\caption{Examples of topographies with 2D translation symmetry. The unit cells of centres (rightmost) for the reference and deformed states are (a) a square,  (b) a rhombus, and (c)  a hexagon. Accordingly, the reference (left) and deformed (middle) tilings have (a) four-fold, (b) three-fold, and (c) six-fold intersections of their interfaces. }
	\label{fig:symmetry}
\end{figure}

Lattices with different symmetries are ubiquitous in nature. Examples include Bravais lattices with translation symmetry, helical structures with helical symmetry, etc. These symmetric lattices can be incorporated into a general framework \textit{objective structures}, firstly demonstrated mathematically in \cite{objective2006james}. Objective structures are constructed by applying a discrete isometry group on a unit cell, which is an atom or a group of atoms. To extend the terminology, the unit cell can be a unit origami structure in order to build \textit{objective origami} with certain symmetry. For example, one can construct Miura origami or helical Miura origami \cite{feng2020helical} by applying the translation group or helical group on a four-fold origami respectively.

Similar ideas apply to the non-isometric origami (or the topography) we study here. We construct three examples in Fig. \ref{fig:symmetry} by applying the 2D translation group on the ``unit cell" \footnote{For a general isometry group $G=\{g_1^p g_2^q: (p, q) \in \mathbb{Z}^2\}$ with $g_1=(\bfQ_1|\bft_1)$ and $g_2 = (\bfQ_2 | \bft_2)$, we follow the operation rules: $g_i(\bfx) = \bfQ_i \bfx + \bft_i$, $g_i^{-1} = (\bfQ_i^T| - \bfQ_i^T \bft_i)$, $g_1 g_2 = (\bfQ_1 \bfQ_2| \bfQ_1 \bft_2 + \bft_1)$, where $i=1,2$, $\bfQ_i \in O(3)$, $\bft_i, \bfx \in \mathbb{R}^3$.}. The unit cell $\calU$ of the reference domain is a square, a rhombus, or a hexagon respectively; see  Fig.~\ref{fig:symmetry}, last of (a) , (b) and (c) respectively. The translation group, $G=\{g_1^p g_2^q: (p, q) \in \mathbb{Z}^2\}$ with $g_1=(\bfI |\bft_1)$ and $g_2 = (\bfI | \bft_2)$, respects the symmetry of the tiling constructed by
\beq
g_1^p g_2^q (\calU).
\eeq
Here $g_1$ and $g_2$ are group generators, $\bfI$ is the identity and $\bft_1, \bft_2 \in \mathbb{R}^2$ are translations consistent with the tiling.
The reference interfaces are straight lines that lead to equal-height deformed centres (recall (\ref{eq:height})). Thus, the translation group $\hat{G}=\{\hat{g}_1^p \hat{g}_2^q: (p, q) \in \mathbb{Z}^2\}$ for the deformed domain is also two dimensional, but with linearly rescaled translations. That is, the group generator $\hat{g}_i=(\hat{\bfI} |\hat{\bft}_i)$ has $\hat{\bft}_i = \lambda ((\bft_i \cdot \bfe_1) \hat{\bfe}_1 + (\bft_i \cdot \bfe_2) \cdot \hat{\bfe}_2)$, for $i=1,2$ and $\hat{\bfI}$ is the $3\times3$ identity. Also recall Fig. \ref{fig:cones}(b) and Eq. (\ref{eq:deformation2}) for the rescaling factor $\lambda$.

We list the unit cells and translation groups in detail:
\begin{enumerate}
	\item[(a).] The four pattern centres in the unit cell are located at $\bfp_1={\bf 0}, \bfp_2 = \bfe_1, \bfp_3 = \bfe_1 + \bfe_2, \bfp_4=\bfe_2$. The generators for the reference domain are $g_1 = (\bfI | \bfe_1)$ and $g_2=(\bfI | \bfe_2)$. The generators for the deformed domain are $\hat{g}_1 = (\hat{\bfI} | \lambda \hat{\bfe}_1)$ and $\hat{g}_2= (\hat{\bfI} | \lambda \hat{\bfe}_2)$.
	\item[(b).] The four centres in the unit cell are located at $\bfp_1={\bf 0}, \bfp_2 = \bfe_1, \bfp_3 =1/2 \bfe_1 +\sqrt{3}/2 \bfe_2, \bfp_4=-1/2 \bfe_1 + \sqrt{3}/2 \bfe_2$. The generators for the reference domain are $g_1 = (\bfI | \bfe_1)$ and $g_2=(\bfI | -1/2 \bfe_1 + \sqrt{3}/2 \bfe_2)$. The generators for the deformed domain are $\hat{g}_1 = (\hat{\bfI} | \lambda \hat{\bfe}_1)$ and $\hat{g}_2= (\hat{\bfI} | \lambda (-1/2 \hat{\bfe}_1 + \sqrt{3}/2 \hat{\bfe}_2))$.
	\item[(c).] The six centres in the unit cell are located at $\bfp_i = \bfR(\frac{i \pi}{6} )\bfe_1, i =1,2,\dots,6$. The generators for the reference domain are $g_1 = (\bfI | 3/2\bfe_1 + \sqrt{3}/2 \bfe_2)$ and $g_2=(\bfI | \sqrt{3} \bfe_2)$. The generators for the deformed domain are $\hat{g}_1 = (\hat{\bfI} | \lambda(3/2\hat{\bfe}_1 + \sqrt{3}/2 \hat{\bfe}_2))$ and $\hat{g}_2= (\hat{\bfI} | \lambda \sqrt{3}\hat{\bfe}_2)$.
\end{enumerate}

These three types of symmetric tilings have four-fold, three-fold, and six-fold intersections. On the deformed domain, the tips (centres of circular patterns) for a tiling have the same height and therefore are on the same plane. We know that the triangles, squares and hexagons are the only three types of regular tessellations. The three cases in Fig. \ref{fig:symmetry} are candidates for the design of active lifters. Among them, only the four-fold tiling has been investigated experimentally \cite{ware2015voxelated, white2015programmable}. In experiment, we observe that
%the inversion of the
tips provide the force to lift the load. Therefore, it is reasonable to conjecture that the density of tips is crucial to the lifter's performance evaluated by the maximal affordable load per unit area. For this reason, the three-fold tiling [Fig. \ref{fig:symmetry}(b)] might have the best performance if we fix the distance between two nearest centres.

\subsubsection{Complex topographies and the inverse design of pixels}
\begin{figure}[!th]
	\centering
	\includegraphics[width=\columnwidth]{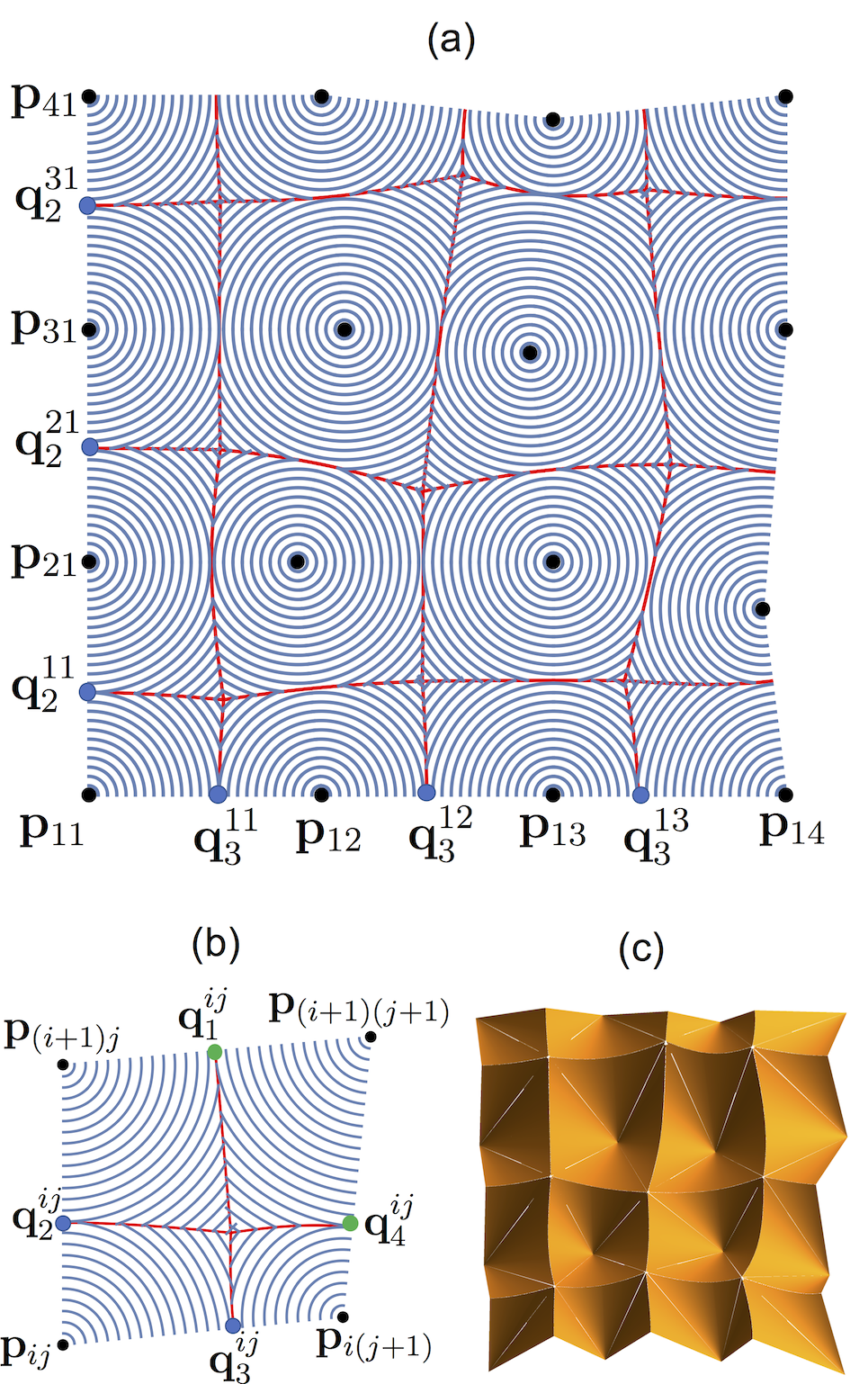}
	\caption{(a) An example of irregular four-fold tiling on the reference domain. (b) The ij-th unit cell of the tiling. The four centres  and two input offsets $\bfq_2^{ij}, \bfq_3^{ij}$ (blue dots) will determine the output offsets $\bfq_1^{ij}, \bfq_4^{ij}$ (green dot). (c) The deformed irregular four-fold tiling.}
	\label{fig:4fold}
\end{figure}
\begin{figure}[!th]
	\centering
	\includegraphics[width=\columnwidth]{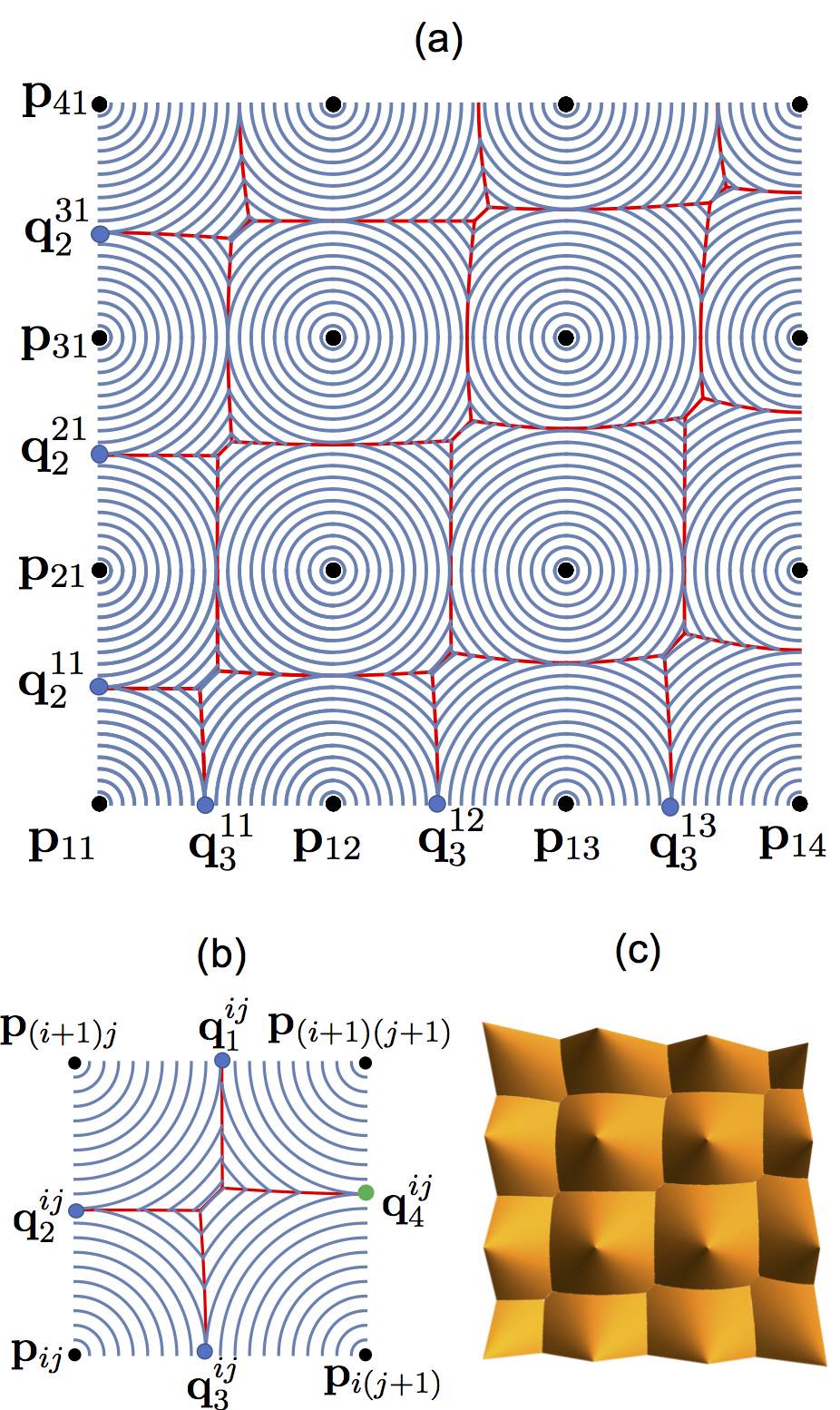}
	\caption{(a) An example of irregular three-fold tiling on the reference domain. (b) The ij-th unit cell. The four centres  and three input offsets $\bfq_1^{ij}, \bfq_2^{ij}, \bfq_3^{ij}$ (blue dots) will determine the output offset $\bfq_4^{ij}$ (green dot). (c) The deformed irregular three-fold tiling.}
	\label{fig:3fold}
\end{figure}
In this section, we design more complex tilings using three-fold and four-fold intersections with no translation symmetries. A tiling usually differs from a single intersection in that it requires more global restrictions $-$  we have to arrange different unit cells in a compatible way. In origami community,
the global restriction, called {\it global compatibility}, is the key idea to design tilings like rigidly and flat-foldable origami \cite{lang2018rigidly,feng2020designs}. The global compatibility is automatically satisfied for symmetric patterns in Fig.~\ref{fig:symmetry}, whereas  irregular patterns have to obey different design principles.

Fig. \ref{fig:4fold} shows an example of irregular four-fold tiling. To design the tiling, we prescribe all the centres $\bfp_{ij},~ i\in \{1,\dots,m\},~j\in\{1,\dots,n\}$, and the bottom/left boundary offsets $\bfq_3^{11},\dots, \bfq_3^{1(n-1)}, \bfq_2^{11}, \dots, \bfq_2^{(m-1)1}$, as shown in Fig. \ref{fig:4fold}(a). Recall that for a single four-fold intersection, the four centres and two given offsets will determine the other two offsets. Specifically, in the ij-th unit cell shown in Fig. \ref{fig:4fold}(b), the output offsets $\bfq_1^{ij}$ and $\bfq_4^{ij}$ (dots in green) are determined by the four centres $\bfp_{ij}, \bfp_{i(j+1)}, \bfp_{(i+1)(j+1)}, \bfp_{(i+1)j}$ and two input offsets $\bfq_1^{ij}, \bfq_4^{ij}$ (dots in blue). This gives a marching algorithm for computing the rest offsets and the entire tiling by
\beq
\bfq_2^{i(j+1)} = \bfq_4^{ij},~ \bfq_3^{i(j+1)} = \bfq_1^{(i-1)(j+1)},
\eeq
i.e., the input offsets of the current unit cell are inherited from the output offsets of previous left and bottom unit cells until one reaches the left/bottom boundaries.
Thus,  the entire four-fold tiling is determined by all the centres and by the left/bottom boundary offsets. The result is parallel with  previous work \cite{feng2020designs} of one of us for the design of rigidly and flat-foldable quadrilateral mesh origami, in which the geometrical data on the left/bottom boundaries determine the entire origami pattern.

In comparison with the four-fold tiling, the three-fold tiling [Fig. \ref{fig:3fold}] has more degrees of freedom. In the ij-th unit cell [Fig. \ref{fig:3fold}(b)],  the offset $\bfq_4^{ij}$ is determined by the four centres $\bfp_{ij}, \bfp_{i(j+1)}, \bfp_{(i+1)(j+1)}, \bfp_{(i+1)j}$ and three offsets $\bfq_1^{ij}, \bfq_2^{ij}, \bfq_3^{ij}$, according to Sect. \ref{sec:splitting}. Thus, each unit cell of the three-fold tiling has one additional degree of freedom to assign $\bfq_1^{ij}$. The marching algorithm for the three-fold tiling is similar to the one for the four-fold tiling, but one needs to assign the additional DOF $\bfq_1^{ij}$ for each unit cell.
Recall that the relative height between the deformed $\bfp_{(i+1)j}$ and $\bfp_{(i+1)(j+1)}$ is determined by the position of $\bfq_1^{ij}$.
Then, the extra DOF of $\bfq_1^{ij}$ in the ij-th unit cell provides the opportunity of designing the relative heights. Since every unit cell has one free $\bfq_1^{ij}$ to assign, technically one can design the height of every inner tip of the tiling.
Here inner tips mean all the tips of the tiling except for those on the left/bottom boundary ($\bfp_{i1}$ and $\bfp_{1j}$, $ i\in \{1,\dots,m\},~ j\in\{1,\dots,n\}$).

\begin{figure}[!h]
	\centering
	\includegraphics[width=0.8\columnwidth]{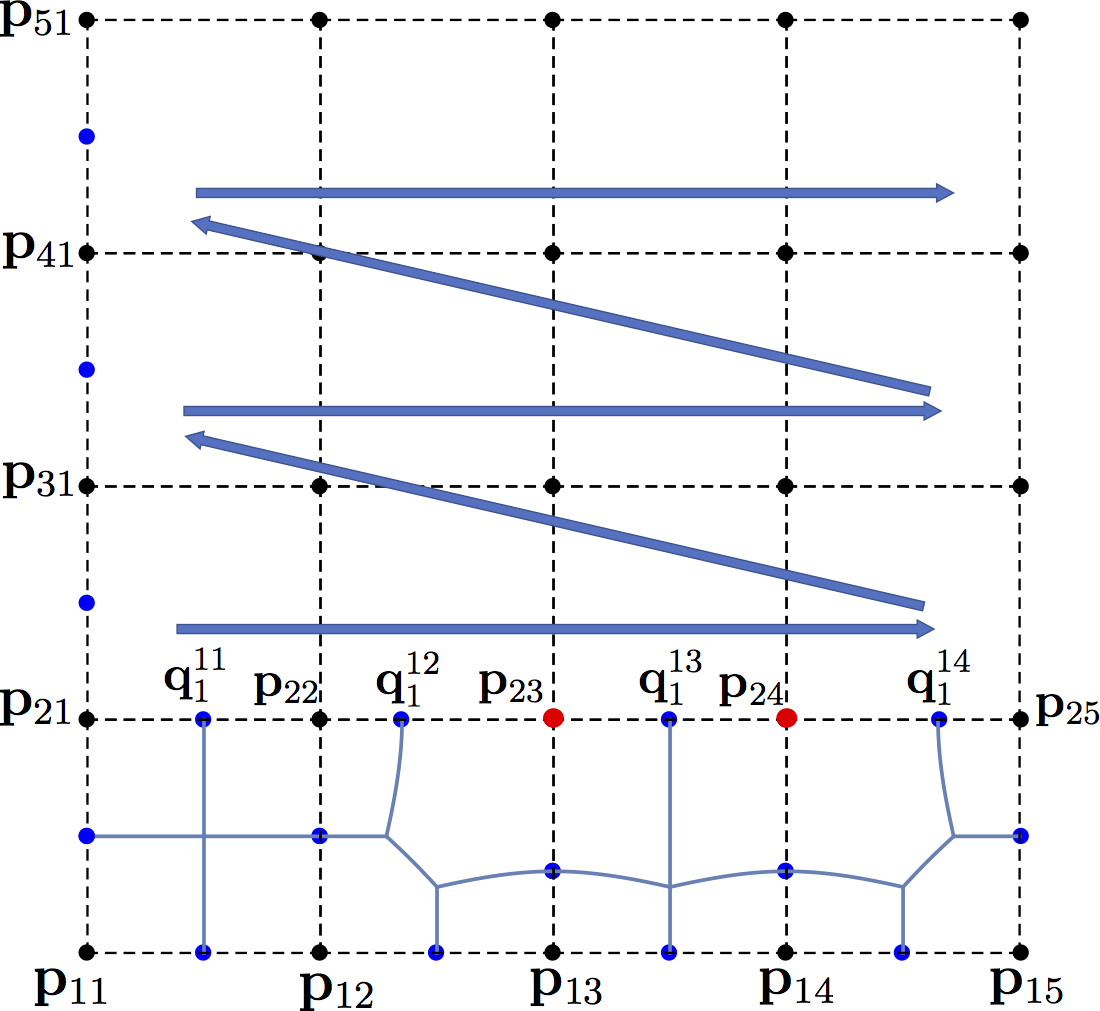}
	\caption{The schematic of inverse design. The centres (red and black dots) form a square lattice and the boundary offsets (blue dots) are bisectors. The red dots are designed to be higher than the black dots in the deformed domain by moving $\bfq_1^{ij}$ properly. The positions of $\bfq_1^{ij}$ are determined successively by the target image, from left to right, from bottom to top, as the blue arrows indicate -- see text. }
	\label{fig:sketch}
\end{figure}

In the end, relying on the property of 3-fold tiling, we highlight the design of arrays of pixels that have the basic functionality of displaying monochrome images. Basically, we design a three-fold tiling consisting of circular patterns and compatible interfaces on the reference domain and then actuate it with heat or illumination.
The tips with larger heights on the deformed domain will act as pixels and display the target image, whereas the others act as the background.
We introduce the design strategy by following the schematic in Fig. \ref{fig:sketch}. We start with a square lattice of centres $\bfp_{ij}$ as the array of pixels. The left/bottom boundary offsets are pre-assigned as bisectors. After deformation,  centres $\bfp_{ij}$ are in one of  the two states, with larger heights denoted by $f(\bfp_{ij}) = 1$ or smaller heights denoted by $f(\bfp_{ij}) = 0$. The value of $f(\bfp_{ij})$ is determined by the target image. Recall that if the centres and left/bottom boundary offsets are given, each two connected three-fold unit cell has one additional free parameter $\bfq_1^{ij}$ to assign. According to the values of $f(\bfp_{ij})$, the position of $\bfq_1^{ij}$
is defined by
\begin{widetext}
\begin{align}
	\bfq_1^{ij} = \begin{cases}
		(1-\delta) \bfp_{(i+1)j} + \delta \bfp_{(i+1)(j+1)},~ &\text{if}~f(\bfp_{(i+1)(j+1)} ) - f(\bfp_{(i+1)j}) = 1 \\
		1/2, ~&\text{if} ~f(\bfp_{(i+1)(j+1)} ) - f(\bfp_{(i+1)j}) = 0\\
		\delta \bfp_{(i+1)j} +(1- \delta) \bfp_{(i+1)(j+1)},~ &\text{if}~f(\bfp_{(i+1)(j+1)} ) - f(\bfp_{(i+1)j}) = -1 \\
	\end{cases}. \label{eq:offset}
\end{align}
\end{widetext}
Here we choose $\delta$ in $(1/2,1)$ to satisfy the sign of $f(\bfp_{ij})$ consistently.  Specifically, we set $\delta = 0.65$ in our examples. Recalling the height difference given by (\ref{eq:height}), the height difference in terms of the size $d$ of pixel and the position $\delta$ of offset  is
\beq
\Delta h = (2\delta - 1) d \sqrt{\lambda^{-2\nu} - \lambda^2},
\eeq
where $d = |\bfp_{11}-\bfp_{12}|$ is the size of pixel on the reference domain.
 For the example in Fig. \ref{fig:sketch}, the red dots are supposed to have larger heights on the deformed domain.
 These offsets $\bfq_1^{ij}$ are computed by (\ref{eq:offset}) successively, from left to right, from bottom to top. The entire tiling is then determined.
\begin{figure}[!th]
	\centering
	\includegraphics[width=\columnwidth]{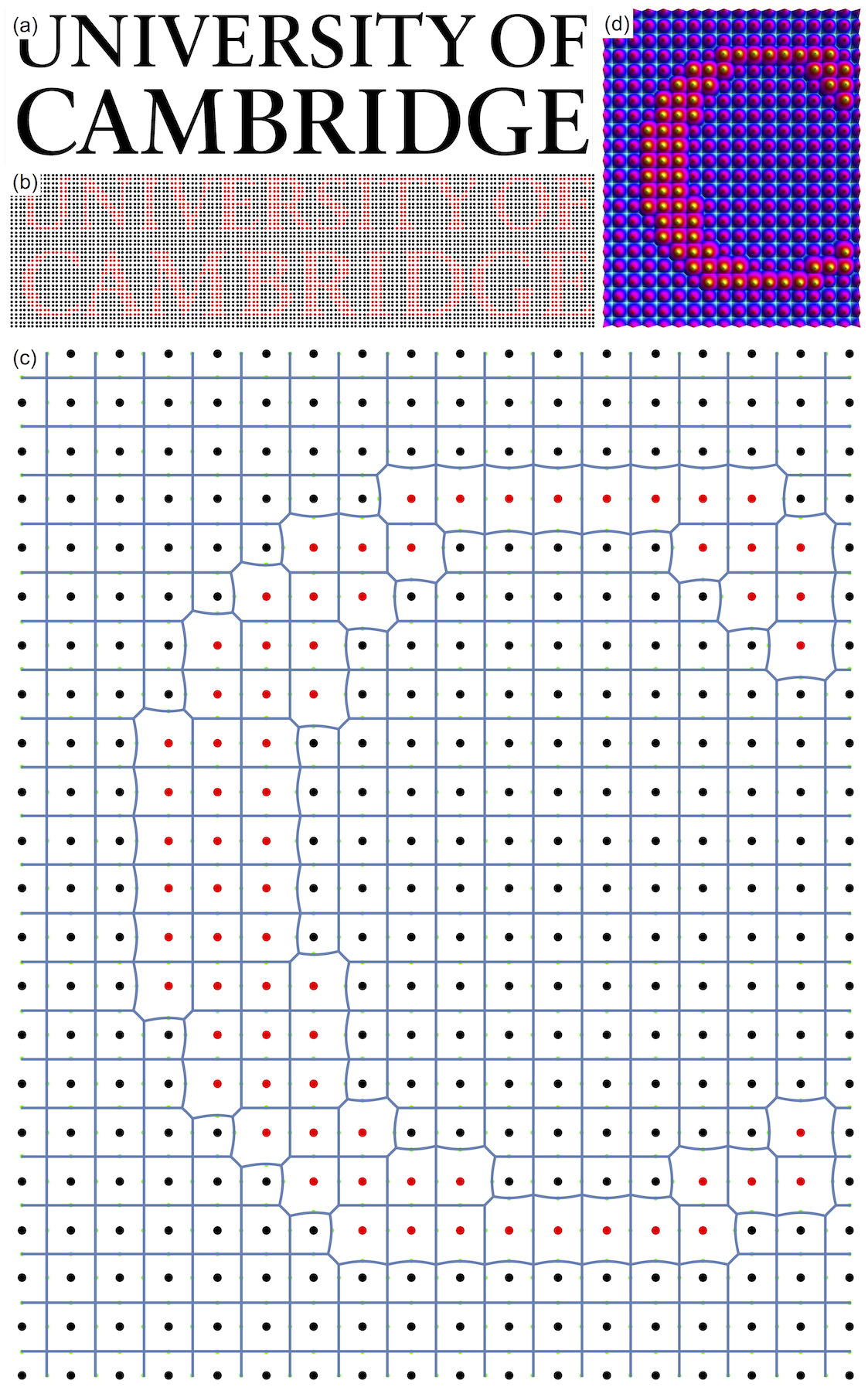}
	\caption{(a) Target image\footnote{See \href{https://www.cam.ac.uk/}{https://www.cam.ac.uk/ } for the symbol ``UNIVERSITY OF CAMBRIDGE".}. (b) 142$\times$37 pixels  designed to illustrate the target image. The red dots represent the centres of circular patterns that are supposed to have larger heights than the black dots in the deformed domain. (c) The design of ``C" on the reference domain with light blue curves as the metric-compatible interfaces. The reference domain contains four-fold and Type I/Type II connected three-fold intersections. (d) The deformed domain displays ``C" as we expect. The contour colors represent the heights.}
	\label{fig:inverse}
\end{figure}

In principle,  our design can display any monochrome image. We end up the discussion by presenting an example of display showing the symbol ``UNIVERSITY OF CAMBRIDGE" [Fig. \ref{fig:inverse}]. The target image [Fig. \ref{fig:inverse}(a)] is discretized as an array of $142\times37$ pixels [Fig. \ref{fig:inverse}(b)], with dots representing centres of circular patterns. The red dots (tips) have larger heights than the black dots in the deformed domain. Following the design principle, we explicitly present the design of ``C" in Figs. \ref{fig:inverse}(c)(d) with the reference and deformed domains. The reference domain [Fig. \ref{fig:inverse}(c)] consists of four-fold and Type I/Type II  connected three-fold intersections, with light blue curves as the metric-compatible interfaces. The deformed domain, equipped with the contour color of heights, displays the ``C" correctly as we expect.

\section{Discussion}
%\vspace{-.4cm}
In this work, we have presented a thorough idea of designing complex topographies and non-isometric curved fold origami using LCE films with circular director patterns. We employ a metric compatibility condition generalized to study the curved interfaces in both the reference and deformed domains. We have chosen to focus our attention on  patterns with hyperbolic metric-compatible interfaces, due to the promising property that the deformed interfaces can be matched by translations. Symmetric patterns, including square, triangular and hexagonal cases,  complement existing designs of active load lifters. Irregular patterns with three-fold and four-fold intersections provide tremendous flexibility for inverse design, for example, the design of pixels to display target images. In comparison with the 3-D director pattern at thin film limit by Plucinsky \cite{plucinsky2018actuation}
 and the pure 2-D director field by Griniasty \cite{griniasty2019curved}, our design seems more robust, due to the robustness of cones, even though the realisation is limited to discretized images.

 Relaxation of actuated structures via isometries to reduce bend energy is to be expected, and  the  less constrained cases we describe will produce less sharp features than predicted. An example would be an isolated concentric square director field, rather than concentric circles. A simple actuation would be to a square pyramid, by the logic of Fig.~\ref{fig:cones}. But the 4 creases leading from the base to the tip can have their bend energy reduced if there is relaxation to a circular cone, albeit one not with a simple circle at the bottom of its skirts. On the other hand, a square array of such concentric square director fields cannot relax to circular cones, at least close to the bases that remain in a square array, even if the individual tips lose some of their square pyramid character. Differently, the lines of R-1 connectedness between neighbouring, e.g. circular, actuated patterns can relax bend energy. For instance Ware \textit{et al} \cite{ware2015voxelated}, Fig. S5 in their supplementary material, shows a 3$\times$1 strip of connected actuated cones relaxing its bends along the R-1 connected lines to a more diffuse form. We return to finite element analysis of the actuated building blocks and complex structures that we have discussed above.

Also to within differences of isometries, and hence bend energies, are various possible alternative actuated landscapes belonging to the same director pattern. Examples are found already in the 3$\times$1 strips of cones of Ware \textit{et al} \cite{ware2015voxelated}, Fig. 3, where either all three cones can pop the same way, or the middle or an end cone can pop in the opposite sense from the other two. The bend energies are different in each case. Even within the two 2 + 1 possibilities, there are different ways the R-1 connected boundary can buckle in the activated state -- see Fig. S4(B) of their supplementary material. The arrays of cones that are loaded in order to actuate as lifters do not seem to suffer these complex breaking of symmetries. Again, we return to this question in a fuller analysis, including FEA.

On the theoretical front, the idea of metric-compatible interfaces between different smooth director patterns can be quite general.  For example, the director patterns can be spirals  as we mentioned in the beginning. Then the deformed interfaces will be more complex due to the complex deformations induced by these patterns. The distribution of GC associated with a general director pattern having interfaces and defects will also be challenging to identify. For these scenarios, more effort from both theoretical and computational sides are needed. For practical applications, it would be particularly valuable for  LCE experiment to realise and test the complex topographies based on our design principle, to achieve optimal actuators, LCE pixels, and soft robotics.

\textbf{Acknowledgment.}
FF and MW were supported by the EPSRC [grant number    EP/P034616/1].
JSB was supported by a UKRI ``future leaders fellowship''  [grant number MR/S017186/1].

\bibliographystyle{ieeetr}
\bibliography{evolving}
\end{document}